\documentclass[aps,prb,superscriptaddress,showkeys]{revtex4-2}
\usepackage[utf8]{inputenc}
\usepackage{amsmath,amssymb,amsfonts}
\usepackage{multirow}
\usepackage{graphicx,float}
\usepackage[colorlinks=true,
citecolor=blue,
linkcolor=blue,
urlcolor=blue]{hyperref}
\usepackage{footmisc}
\begin{document}

\title{Strain Driven Anomalous Anisotropic Enhancement in the Thermoelectric Performance of  Monolayer MoS$_{2}$}
\author{Saumen Chaudhuri}
\affiliation{Department of Physics, Indian Institute of Technology Kharagpur, Kharagpur 721302, India}
\author{Amrita Bhattacharya}
\affiliation{Department of Metallurgical Engineering and Materials Science, IIT Bombay, Mumbai 400076, India}
\author{{A. K. Das}}
\affiliation{Department of Physics, Indian Institute of Technology Kharagpur, Kharagpur 721302, India}
\author{{G. P. Das}}
\affiliation{Research Institute for Sustainable Energy, TCG Centres for Research and Education in Science and Technology, Sector V, Salt Lake, Kolkata 700091, India}
\author{B. N. Dev}
%\thanks{corresponding author}
\email[corresponding author: ]{bhupen.dev@gmail.com}
\affiliation{Department of Physics and School of Nano Science and Technology, Indian Institute of Technology Kharagpur, Kharagpur, 721302, India}
\affiliation{Centre for Quantum Engineering, Research and Education, TCG Centres for Research and Education in Science and Technology, Sector V, Salt Lake, Kolkata 700091, India}

\begin{abstract}
First principles density functional theory based calculations have been performed to investigate the strain and temperature induced tunability of the thermoelectric properties of monolayer (ML) MoS$_2$. Modifications in the electronic and phononic transport properties, under two anisotropic uniaxial strains along the armchair (AC) and zigzag (ZZ) directions, have been explored in detail. Considering the intrinsic carrier-phonon scattering, we found that the charge carrier mobility ($\mu$) and relaxation time ($\tau$) increase remarkably for strains along the ZZ direction. Concomitantly, strain along the ZZ direction significantly reduces the lattice thermal conductivity ($\kappa_\text{L}$) of ML-MoS$_2$. The combined effect of shortened phonon relaxation time and group velocity, and the reduced Debye temperature is found to be the driving force behind the lowering of $\kappa_\text{L}$. The large reduction in $\kappa_\text{L}$ and increase in $\tau$, associated with the strains along the ZZ direction, act in unison to result in enhanced efficiency and hence, improved thermoelectric performance. Nearly $150\%$ enhancement in the thermoelectric efficiency can be achieved with the optimal doping concentration. We, therefore, highlight the significance of in-plane tensile strains, in general, and strains along the ZZ direction, in particular, in improving the thermoelectric performance of ML-MoS$_2$. 
\end{abstract}


\date{\today}
\maketitle

\section{Introduction}
Development and large-scale deployment of clean and green energy resources have become a quintessential challenge for mankind \cite{owusu2016review}. Thermoelectric generators (TEGs), based on the principle of Seebeck effect \cite{zhang2015thermoelectric, freer2020realising}, can convert a temperature difference to useful voltage and thus, are ideally suitable for waste heat recovery \cite{orr2017prospects}. However, the poor efficiency of the TEGs poses the main challenge of thermoelectric research \cite{liu2015current}. The thermoelectric device efficiency is primarily dependent on the intrinsic transport properties of constituent thermoelectric materials (TEMs), which is given by the dimensionless quantity ZT, which is also called the figure of merit. $\text{ZT}=  \frac{\text{S}^{2}\sigma}{\kappa} \text{T}$, where $\sigma$ is the electrical conductivity, S is the Seebeck coefficient or the thermopower, $\kappa = \kappa_\text{e} + \kappa_\text{L}$ is the total thermal conductivity, which is the sum of electrical ($\kappa_\text{e}$) and lattice thermal conductivity ($\kappa_\text{L}$), at a given temperature T. 
    
To date, the highest ZT values in typical TEMs are reported to be in the range of 2.5 to 3 \cite{patel2020bulk, beretta2019thermoelectrics, zhao2014ultralow, nshimyimana2020discordant} without nanostructuring. Tuning the ZT of a given TEM is an extremely tedious task, since all the electronic transport coefficients viz. the S, $\sigma$ and $\kappa_\text{e}$ are dependent on the electronic carrier concentration (\emph{n}) and therefore, can not be tuned individually. However, $\kappa_\text{L}$ is the only quantity which does not directly depend on \emph{n}. Hence, the two main routes to maximize the ZT are (a) band engineering to maximize the power factor \cite{bilc2015low, pei2011convergence} and (b) phonon engineering for lowering the $\kappa_\text{L}$ of the material \cite{dresselhaus2007new, zhang2022reduced, gautam2020enhanced, li2012vacancy, ding2015manipulating}.  

Due to the low-dimensional structure and the resulting interface phonon scattering and phonon confinement effects, the reduction in $\kappa$ occurs naturally in $2$D materials. With this advantage over their bulk counterparts, various $2$D-materials, viz. transition metal dichalcogenides (TMDCs) (e.g., MoS$_2$, WS$_2$, MoSe$_2$, and HfS$_2$), group IVA-VA compounds (e.g., SnSe, SnS, GeSe) etc., have emerged as potential thermoelectric material with considerably high thermoelectric efficiency \cite{dresselhaus2007new}. Layered transition metal di-chalcogenides (TMDCs) have drawn tremendous interest due to their stable crystal structure, intrinsic semiconducting nature and high charge carrier mobility \cite{wang2012electronics, guo2013theoretical, akinwande2014two, chaudhuri2022strain, radisavljevic2011single}. Substantial research has been carried out in recent times towards the development of single-layer TMDCs both using simulations and experiments. Some of the TMDCs such as WS$_2$ \cite{patel2020high}, WSe$_2$ \cite{kumar2015thermoelectric}, HfS$_2$ \cite{wang2021improved}, SnSe$_2$ \cite{su2013snse2} etc. have shown notably high thermoelectric performance. Apart from that, recently, a novel type of two-dimensional TMDCs, i.e., (AX)$_2$ (A = Si, Ge, Sn, Pb; X = Se, Te), has been predicted theoretically to have high thermoelectric efficiency \cite{jia2022high}. Besides the single-layer 2D materials, van der Waals (vdW) heterostructures combining various TMDCs have led to an upsurge in research interest due to their superior physical and chemical properties beyond the single parent material \cite{jia2022recent}. Due to the strong interfacial phonon scattering and the resulting intrinsically low thermal conductivity, the vdW heterostructures offer significantly high thermoelectric performance. Among the various single-layer TMDCs, MoS$_2$ has been explored extensively as a potential thermoelectric material, both theoretically \cite{bhattacharyya2014effect, huang2014theoretical, wickramaratne2014electronic, jena2017compressive, guo2013high, xiang2019monolayer, jin2015revisit} and experimentally \cite{hippalgaonkar2017high, kayyalha2016gate} and the monolayer (ML) counterpart of MoS$_2$ appears to be a reasonably good thermoelectric material \cite{bhattacharyya2014effect, jena2017compressive, xiang2019monolayer, hippalgaonkar2017high, kayyalha2016gate, jin2015revisit}. From experimental investigations, the power factor ($\text{S}^{2}\sigma$) of ML-MoS$_2$ at room temperature is found to be $8.5 \times 10^{-3}$ $\text{W}\text{K}^{-2} \text{m}^{-1}$, which is the highest among all thermoelectric materials \cite{hippalgaonkar2017high}. However, the ZT value obtained for ML-MoS$_2$ is low and not useful for practical purposes. Infact, a ZT value of only $\sim 0.26$ at $500\,\text{K}$ has been achieved with ML-MoS$_2$ \cite{jin2015revisit}. Although in some theoretical reports, an improved ZT value is achieved \cite{huang2013thermoelectric}, they are well below the values obtained with other TMDCs such as MoSe$_2$ ($0.8$), WSe$_2$ ($0.9$), WS$_2$ ($1.1$), SnSe$_2$ ($2.95$), and HfS$_2$ ($1.09$) [as shown in \cite*{rai2020electronic} and references therein]. The primary reason behind the low thermoelectric efficiency of ML-MoS$_2$ is its rather high lattice thermal conductivity ($\kappa_\text{L}$), upon which the ZT value is inversely dependent. The values of $\kappa_\text{L}$ for monolayer MoS$_2$ at $300\,\text{K}$, found in theory \cite{rai2020electronic} and experiment \cite{yan2014thermal} are $30.1$ and $34.5$ $\text{Wm}^{-1}\text{K}^{-1}$ respectively. These values of $\kappa_\text{L}$ are considerably higher compared to the $\kappa_\text{L}$ values obtained with other TMDCs at the same temperature, such as WSe$_2$ ($3.93$ $\text{Wm}^{-1}\text{K}^{-1}$) ZrS$_2$ ($3.29$ $\text{Wm}^{-1}\text{K}^{-1}$), ZrSe$_2$ ($1.2$ $\text{Wm}^{-1}\text{K}^{-1}$), HfS$_2$ ($5.01$ $\text{Wm}^{-1}\text{K}^{-1}$), and HfSe$_2$ ($1.8$ $\text{Wm}^{-1}\text{K}^{-1}$) [as shown in \cite*{rai2020electronic} and references therein]. The high value of $\kappa_\text{L}$ and the resulting low efficiency hinders the practical application of ML-MoS$_2$ as a thermoelectric material. Recently, Janus TMDC monolayers, with an X-M-Y (M = Mo, W, Hf, Zr, and X/Y = S, Se, Te) sandwich structure have emerged as a plausible solution for future thermoelectric research. The Janus TMDC monolayers possess significantly low lattice thermal conductivity due to their softened phonon dispersion, reduced group velocity and strong phonon anharmonicity \cite{bera2020ultralow, patel2020high, guo2018phonon}. Such features are highly beneficial for thermoelectric applications. For example, owing to the very low lattice thermal conductivity, the Janus WSTe monolayer offers incredibly high thermoelectric efficiency, which is much higher compared to the pure WS$_2$ monolayer \cite{patel2020high}.   

Efforts have been made to improve the thermoelectric efficiency of ML-MoS$_2$. With the application of an external electric field, a significant enhancement in the Seebeck coefficient of ML-MoS$_2$ has been achieved \cite{buscema2013large}. Doping with impurity atoms is one of the most studied approaches in improving the thermoelectric performance of a material \cite{kong2018realizing, gangwar2019ultrahigh, gautam2020enhanced}. However, the introduction of dopants brings a permanent change into the material and sometimes forms a secondary phase as well \cite{kong2018realizing, gangwar2019ultrahigh, gautam2020enhanced}. Strain engineering, on the other hand, can be a potential alternative to tune the electronic and thermoelectric properties, owing to its simplicity and reversibility \cite{das2014microscopic, guzman2014role}. For instance, the thermoelectric performance of monolayer ZrS$_2$ and HfS$_2$ is improved via the application of mechanical strain \cite{lv2016strain, wang2021improved}. For monolayer, few-layer and bulk MoS$_2$ as well, the application of strain has proven to be an efficient way to tune and improve the thermoelectric performance \cite{bhattacharyya2014effect}. There have been efforts to implement various in-plane strains on monolayer and few-layer MoS$_2$ experimentally by growing the films on stretchable polymer \cite{pu2013fabrication, rice2013raman, castellanos2013local, zhu2013strain, conley2013bandgap, tan2018raman} or lattice-mismatched substrates \cite{ji2013epitaxial}. Recent reports have demonstrated that very high values of tensile strains for example, $25\%$ on ML-graphene \cite{lee2008measurement} and up to $11\%$ on ML-MoS$_2$ \cite{bertolazzi2011stretching} can be applied using the nano-indentation technique. On the other hand, the application of compressive strain is always associated with the possibility of ripple formation or warping of the film \cite{jiang2014buckling, li2019molecular}. Thus, from the experimental point of view, the enhancement in the thermoelectric performance is desirable with tensile strains rather than with compressive strains, owing to the convenience of strain application.

ML-MoS$_2$ is a well-studied thermoelectric material and therefore, its performance has been avidly explored by researchers in the past. However, a wide range of values for the transport parameters prevail in literature for ML-MoS$_2$, which makes the understanding of the transport characteristics and the impact of strain on it baffling. Thus, it is essential to explore the transport properties of ML-MoS$_2$ in greater detail. Earlier it was reported that the thermoelectric efficiency of ML-MoS$_2$ can be improved by applying in-plane compressive strain \cite{jena2017compressive}. While this is true, the possible improvement that can be achieved with tensile strain is not thoroughly explored, since the impact of strain on the lattice dynamics and phonon transport is not exhaustively studied. Furthermore, certain important aspects, such as the anisotropy in the thermoelectric performance when subjected to in-plane uniaxial tensile strain, or the possible modifications in the charge carrier mobility when stretched along a certain direction (zig-zag or arm-chair), have not been hitherto reported in the literature. In this work, the electronic and phonon transport properties of ML-MoS$_2$ and its modification with uniaxial tensile strains along the armchair (AC) and zigzag (ZZ) directions have been explored in detail. The direction-dependent enhancement in the thermoelectric efficiency of single-layer MoS$_2$ with in-plane strains is highlighted. The underlying mechanism behind the strain-induced anisotropic modifications in the charge carrier transport and phonon dynamics is unveiled through systematic theoretical calculations.

\section{Computational details}
First-principles calculations have been performed using ab-initio density functional theory (DFT) as implemented in the Vienna Ab Initio Simulation Package (VASP) \cite{kresse1996efficient, kresse1996efficiency} together with projector augmented wave (PAW) potentials to account for the electron-ion interactions \cite{kresse1999ultrasoft}. The electronic exchange and correlation (XC) interactions are addressed within the generalized gradient approximation (GGA) of Perdew-Burke-Ernzerhof (PBE) \cite{perdew1996generalized}. In order to explore the implication of non-local exchange and correlation on the electronic band structure, hybrid HSE06 functional has been used for some of the relevant cases. A vacuum thickness of $20 \text{\AA}$ is used to minimize the interaction between the periodic images of the layers. A well-converged Monkhorst-Pack \cite{monkhorst1976special} k-points set of $21 \times 21 \times 1$ together with a plane wave cutoff energy of $450$ eV are used for the geometry relaxation, and a well-converged, denser k-mesh is used for the post-relaxation calculations. For all the strained structures, while the scaled lattice parameter is kept fixed, the atomic coordinates are allowed to relax until the forces on the atoms are optimized to be less than $0.01$ eV/$\text{\AA}$. 

To calculate the temperature- and doping-level-dependent changes in the thermoelectric parameters such as the Seebeck coefficient (S), electrical conductivity ($\sigma$), power factor ($\text{S}^{2}\sigma$) etc., semi-classical Boltzmann transport theory implemented within the BoltzTraP package \cite{madsen2006boltztrap} is used. The BoltzTraP package calculates the thermoelectric properties under the constant relaxation time approximation (CRTA) and is based on the electronic energy eigenvalues calculated using VASP. To go beyond the CRTA, the charge carrier mobility and relaxation time are calculated explicitly using the AMSET code \cite{ganose2021efficient} taking the intrinsic carrier-phonon scattering effects into consideration. The material-specific inputs such as the elastic constants, dielectric constants and the deformation potential are estimated from first-principle calculations using VASP. The absolute deformation potentials \cite{resta1990absolute} of the valence and the conduction bands in all strained cases have been calculated using the vacuum energy level as a reference.    

The phonon dispersion curves are calculated based on the supercell approach using the finite displacement method implemented in the phonopy package \cite{togo2015first}. The amplitude of the displacements is fixed at $0.015\, \text{\AA}$. The second order harmonic interatomic force constants (IFC), required for the phonon dispersion, are calculated using a convergence-checked $4 \times 4 \times 1$ unit cell together with a strict energy convergence criterion of $10^{-8}$ eV. To investigate the lattice thermal transport, the Boltzmann transport equation (BTE) for phonons is solved under the relaxation time approximation (RTA) as implemented in phono$3$py \cite{togo2015distributions}. The second- and third-order interatomic force constants (IFC) are calculated using a convergence-checked $4 \times 4 \times 1$ and $2 \times 2 \times 1$ supercell based on the relaxed unit cell, respectively. Fourth- and higher-order IFCs are not taken into consideration due to their presumably small contribution to lattice thermal transport. Well-converged k-meshes are used to sample the Brillouin Zone (BZ) of the supercells. To accurately compute the lattice thermal conductivity both in the unstrained and the strained cases, a dense q-mesh of $51 \times 51 \times 1$ is used to sample the reciprocal spaces of the primitive cells. The mode-resolved lattice thermal conductivity ($\kappa_{\text{L}_\lambda}$), phonon group velocity ($\text{v}_\lambda$) and phonon relaxation time ($\tau_\lambda$) are extracted using python-based extensions as implemented in phonopy-VASP. 

Equilibrium molecular dynamics (EMD) calculations have been performed to assess the thermal stability of ML-MoS$_2$ at different temperatures using the LAMMPS package \cite{thompson2022lammps}. The EMD simulations are performed on a sufficiently large supercell of ML-MoS$_2$ imposing in-plane periodic boundary conditions and using the NPT ensemble \cite{martyna1992nose, martyna1994constant} within the Stillinger-Weber potential developed by Wen et al. \cite{wen2017force}. The system underwent a temperature increase to $300$\, K, $600$\, K, $900$\, K, $1700$\, K and $2000$\, K using the Nose-Hoover thermal bath \cite{evans1985nose} and then equilibrated at a particular temperature for $200$\, ps to study the thermal stability.

\section{Results and Discussion}
\subsection{Unstrained monolayer MoS$_2$}

\begin{table}[h]
	\caption{\label{tab:Table 1} Calculated lattice parameters of the hexagonal unit cell and the orthorhombic conventional cell of monolayer MoS$_2$.}
	\begin{tabular*}{0.7\textwidth}{ c @{\extracolsep{\fill}} c c c }
		\hline
		Crystal structure & a (\AA) & b (\AA) & $\gamma$ (degree) \\
		\hline 
		Hexagonal & 3.17 & 3.17 & 120 \\
		\hline  
		Orthorhombic & 3.17 & 5.51 & 90 \\
		\hline    
	\end{tabular*}
\end{table}

\begin{figure}[h!]
	\centering
	\includegraphics[scale=0.45]{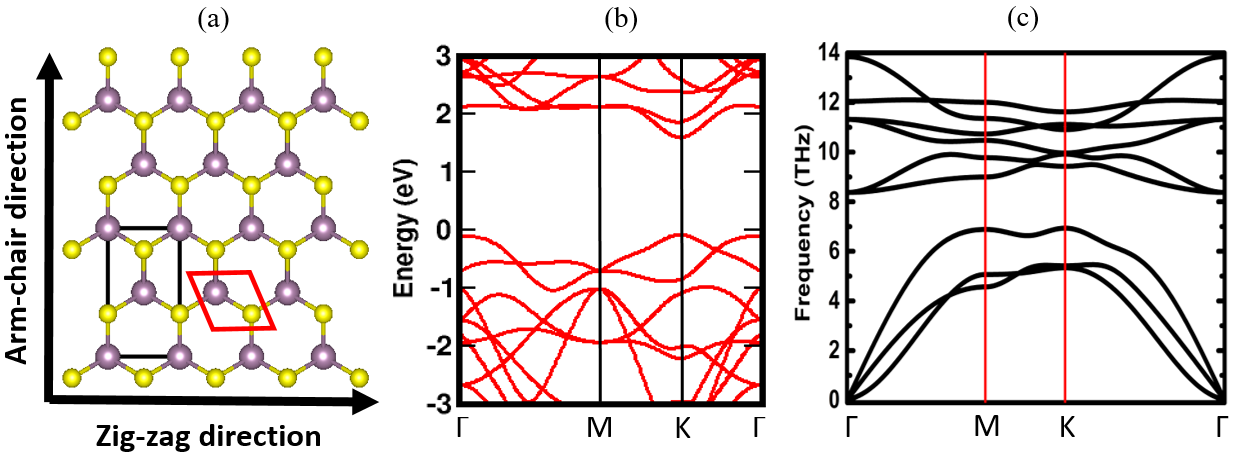}
	\caption{ (a) Crystal structure of monolayer MoS$_2$ (Mo and S atoms are shown in purple and yellow, respectively). The arrows show the arm-chair and the zig-zag directions. The hexagonal unit cell and the orthorhombic conventional cell are shown by the red and black boxes, respectively. (b) Electronic and (c) phonon band structure of unstrained ML-MoS$_2$ plotted along the high symmetry path of the irreducible Brillouin zone (BZ).}
	\label{Fig.1}
\end{figure}

Before going into the details of the strain-induced effects, we first present an overview of the calculated geometrical, electronic and thermoelectric properties of pristine single-layer MoS$_2$ in the 2H phase. In the hexagonal honeycomb crystal structure of ML-MoS$_2$, the Mo and S atomic planes are arranged in a ``sandwich" type of structure (S - Mo - S) with the Mo and S atoms having trigonal prismatic coordination, as shown in Fig. \ref{Fig.1}. In order to apply the mechanical strain along the two non-equivalent lattice directions independently, namely the armchair (AC) and the zigzag (ZZ) directions, an orthorhombic conventional cell (see Fig. \ref{Fig.1}) containing two Mo atoms and four S atoms has been used instead of the hexagonal unit cell. Thus, in the chosen orthorhombic cell the AC and the ZZ crystallographic directions desirably align with the orthogonal lattice vectors. The optimized lattice parameters of both the cells in the unstrained condition are given in Table \ref{tab:Table 1}, which are found to be in good agreement with earlier reports \cite{jena2017compressive, jena2019valley}. 

Unstrained MoS$_2$ in its monolayer (ML) form is found to be semiconducting with both the valence band maximum (VBM) and the conduction band minimum (CBM) at the same high-symmetry point (at K point) of the BZ, as shown in Fig. \ref{Fig.1}(b), resulting in a direct band gap semiconductor with a band gap value of $1.736\, \text{eV}$, which agrees well with earlier reports \cite{chaudhuri2022strain, das2014microscopic, jena2019valley}. The band gap value of unstrained ML-MoS$_2$ from the hybrid functional calculations using the HSE06 functional is found to be $2.184\, \text{eV}$. From Fig. S1 in the supplementary material (see supplementary Sec. A Fig. S1) it can be seen that the electronic band structure of ML-MoS$_2$ calculated using the GGA-PBE functional and HSE06 functional are nearly identical in terms of the band curvature at the extrema, however the implementation of the hybrid functional increases the band gap by $0.448\, \text{eV}$. Since the electronic band dispersions are well produced using the computationally less expensive GGA-PBE functional, we chose to use the GGA-PBE functional in all further electronic structure calculations. The detailed partial density of states (DOS) and the orbital projected band structure (BS) are provided as supplementary material (see supplementary Sec. A Fig. S2). From the DOS and orbital projected BS, it is seen that both the valence band and conduction band edges are primarily composed of Mo-$4d$ and S-$3p$ orbitals, albeit mainly Mo-$4d$ orbitals \cite{chaudhuri2022strain}. The phonon band structure of unstrained single-layer MoS$_2$ is also shown in Fig. \ref{Fig.1}, whereby the absence of imaginary phonon modes throughout the BZ confirms the vibrational stability of the crystal structure. The phonon band structure consists of three low-energy acoustic branches and six optical branches, and they are well separated by a frequency difference of $1.38$ THz  ($46$ cm$^{-1}$). The frequency of the two Raman active modes of ML-MoS$_2$, namely E$_\text{2g}$ and A$_\text{1g}$, are found to be $11.32$ THz ($377\, \text{cm}^{-1}$) and $12.04$ THz ($401\, \text{cm}^{-1}$), respectively. These frequency values are in good agreement with previously reported values \cite{mohapatra2016strictly, qian2014quantum, li2012ideal}.

\begin{figure}[h!]
	\centering
	\includegraphics[scale=0.5]{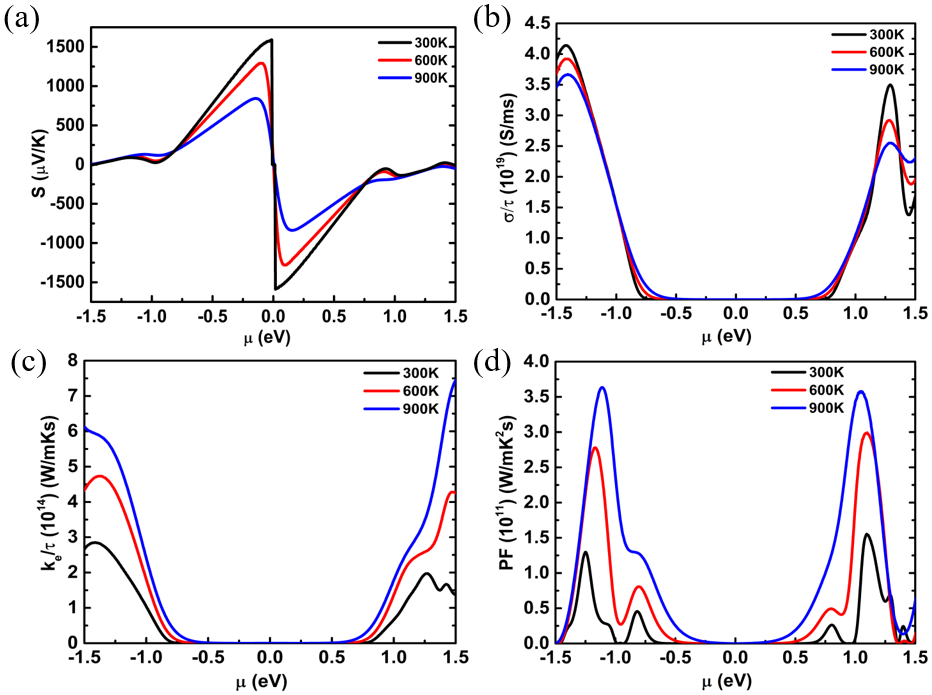}
	\caption{Variation in the thermoelectric parameters of unstrained ML-MoS$_2$ viz., (a) Seebeck coefficient (S), (b) electrical conductivity ($\sigma/\tau$), (c) electronic thermal conductivity ($\kappa_e/\tau$), and (d) power factor, PF ($\text{S}^{2}\sigma/\tau$) plotted as a function of chemical potential ($\mu$) at different temperatures (i.e. at $300$\, K, $600$\, K and $900$\, K). $\mu > 0$ eV corresponds to the n-type carriers or higher electron concentration, and $\mu < 0$ eV corresponds to the p-type carriers or higher hole concentration.}
	\label{Fig.2}
\end{figure}

The thermoelectric transport parameters of unstrained ML-MoS$_2$, i.e. the Seebeck coefficient (S), relaxation time-scaled electrical conductivity ($\sigma/\tau$), electronic thermal conductivity ($\kappa_\text{e}/\tau$) and thermoelectric power factor, PF ($\text{S}^{2}\sigma/\tau$) have been calculated with respect to chemical potential ($\mu$) at different temperatures (T) and are shown in Fig. \ref{Fig.2}. The positive ($\mu > 0$) and the negative values ($\mu < 0$) of the chemical potential stand for the n-doping and p-doping regions, respectively. ML-MoS$_2$ in the unstrained condition exhibits a high S of $1590 \, \mu\text{VK}^{-1}$ at $\text{T} = 300 \,\text{K}$ with optimal doping. The variation in the Seebeck coefficient (S) as a function of chemical potential ($\mu$) at a particular temperature is found to have saw-tooth nature, which is typical of any semiconducting material. The peak value of the relaxation time-scaled thermoelectric PF at $\text{T} = 300 \,\text{K}$ is found to be $15.5 \times 10^{10}\, \text{Wm}^{-1} \text{K}^{-2} \text{s}^{-1}$ in the n-doping region, which is in good agreement with earlier reports using the same exchange-correlation functional \cite{jena2017compressive, rai2020electronic}. The variation in the thermoelectric coefficients with temperature has been presented at $\text{T} = 300\, \text{K}, 600\, \text{K}\, \text{and}\, 900\, \text{K}$. With an increase in T, the peak value of the S and $\sigma/\tau$ decreases. In contrast, the peak value of the PF increases, as shown in Fig. \ref{Fig.2}, which reaches  $36.5 \times 10^{10}\, \text{Wm}^{-1} \text{K}^{-2} \text{s}^{-1}$ at $\text{T} = 900\, \text{K}$. This is comparably as high as some of the best-known thermoelectric materials \cite{wang2021improved, bera2019strain}. Over the years various 2D materials have been investigated as a potential thermoelectric material and the maximum attainable power factor (PF$_\text{max}$) obtained at room temperature has been reported to be $10$ $\times 10^{10}$, $25$ $\times 10^{10}$ and $4$ $\times 10^{10}\, \text{Wm}^{-1} \text{K}^{-2} \text{s}^{-1}$ with WS$_2$ \cite{bera2019strain}, HfS$_2$ \cite{wang2021improved} and PtSe$_2$ \cite{guo2016biaxial}, respectively. Comparing the PF$_\text{max}$ alone, ML-MoS$_2$ stands as a promising thermoelectric material with high potential.  In order to calculate the thermoelectric figure of merit ZT, the variation in the $\kappa_\text{L}$ of ML-MoS$_2$ with temperature has also been investigated. The value of $\kappa_\text{L}$ for ML-MoS$_2$ nanosheet at $300\, \text{K}$ is found to be $24.28\, \text{Wm}^{-1}\text{K}^{-1}$, which is consistent with previously reported theoretical (using MD or DFT framework) \cite{rai2020electronic, cai2014lattice} and experimental \cite{yan2014thermal} values. Pertaining to the increase in scattering, the $\kappa_\text{L}$ decreases sharply with an increase in temperature and reduces to $7.5\, \text{Wm}^{-1}\text{K}^{-1}$ at $900\, \text{K}$. A similar decreasing trend in $\kappa_\text{L}$ has also been observed in other 2D TMDCs such as WS$_2$, WSe$_2$ \cite{mobaraki2018validation} and HfS$_2$ \cite{wang2021improved}. 

Since the transport properties of ML-MoS$_2$ are studied up to a fairly high temperature of $900$\, K, it is necessary to address the thermal stability of the structure and the possible impact of temperature on the electronic properties. To investigate the thermal stability of ML-MoS$_2$, molecular dynamics (MD) calculations have been performed at different temperatures up to $2000$\, K. From the energy plots shown in Fig. S3 (see supplementary Sec. A Fig. S3), it can be seen that the total and the average energy remains stable with obvious temperature induced fluctuations within the simulation time scale for T = $300$\, K, $600$\, K and $900$\, K. However, when a high enough temperature is applied, such as say $2000$\, K, the total energy is found to decrease monotonically with simulation time, indicating the temperature induced instability of the structure. Also, to understand the evolution of the geometry of ML-MoS$_2$ sheet with the application of temperature, the pair correlation function (PCF) has been calculated as a function of temperature. The PCF represents the probability of finding a pair of atoms at a certain separation distance i.e., the bond length of the constituent atoms (see supplementary Sec. A Fig. S3). Distinct and sharp peaks in the PCF indicate that the perfect crystalline order is maintained up to $900$\, K and thus, the structure is stable. However, at $1700$\, K and $2000$\, K the PCF peaks are diffused and the probability of finding a pair of Mo-S atoms is non-zero at all values of separation distance, which suggests the melting or breakdown of the crystalline order of ML-MoS$_2$. Thus, it is safe to assume that ML-MoS$_2$ remains stable at least up to $900$\, K, which is the highest temperature used in the calculation of transport properties. 

To find out an estimate of the effect of temperature on the electronic properties, the linear thermal expansion coefficient (TEC) of ML-MoS$_2$ is calculated within the quasi-harmonic approximation (qha). Details of the calculation of TEC are provided in the supplementary information and the variation in TEC with temperature is shown in Fig. S4 (see supplementary Sec. A Fig. S4). The TEC of ML-MoS$_2$ at $300$\, K and $900$\, K are found to be $6.3 \times 10^{-6}$ /K and $7.2 \times 10^{-6}$ /K, respectively, which are in good agreement with earlier reports \cite{sevik2014assessment, huang2014correlation}. Due to the low values of TEC, it can be assumed that even at the highest temperature considered in the present work ($900$\, K) the lattice expansion and the resulting impact on the electronic band structure will be insignificant. It is also reported that the thermal shifts in the Raman modes at high temperatures are primarily caused by the multi-phonon scattering and thermal expansion plays a negligibly small role. The reported values of the shift rates (cm$^{-1}$/K) of the E$_2g$ and A$_1g$ Raman modes of ML-MoS$_2$ caused by the thermal expansion are $-0.0019$ and $-0.0010$, respectively \cite{huang2014correlation}. It is, therefore, legitimate to neglect the temperature induced effects on the electronic and phononic band structures of ML-MoS$_2$.

\subsection{Effect of Strain}
\subsubsection{Effect of strain on electronic properties}

\begin{figure}[h!]
	\centering
	\includegraphics[scale=0.4]{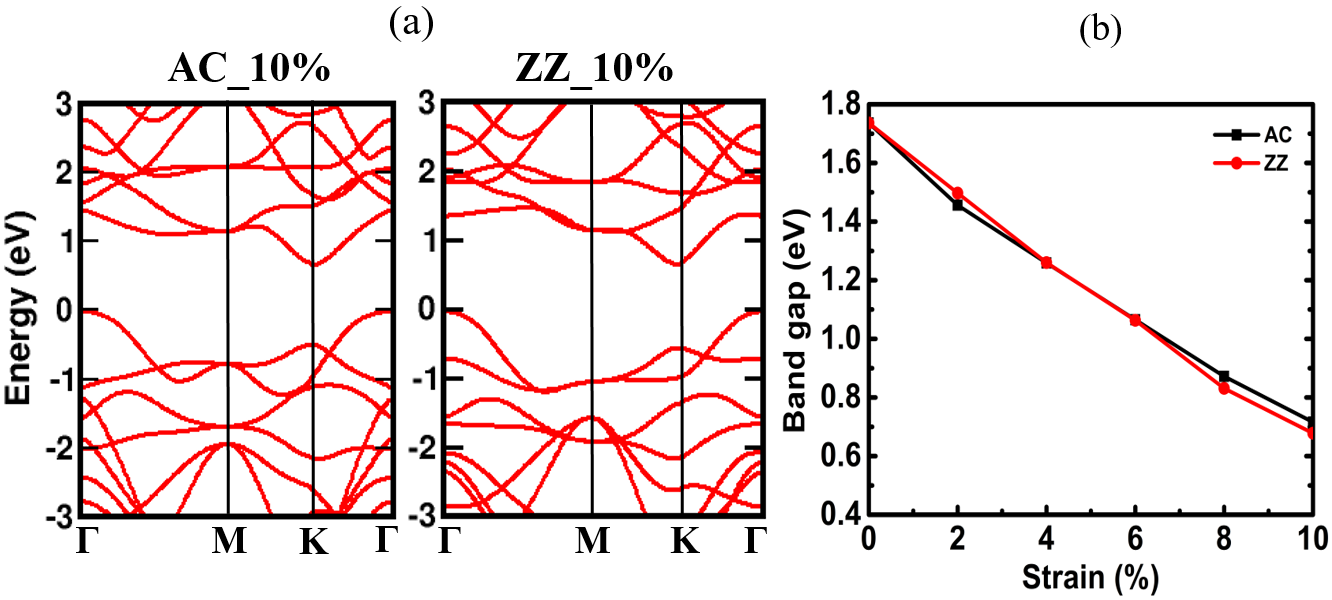}
	\caption{(a) Band structure plots of ML-MoS$_2$ in the extreme strained ($10\, \%$) conditions and (b) variation in the band gap under the uniaxial strain along the AC and the ZZ directions. The strain profile is written at the top of corresponding band structure plots.}
	\label{Fig.3}
\end{figure}

To elucidate the importance of strain engineering in tuning the electronic properties of single-layer MoS$_2$, a series of in-plane uniaxial tensile strains along different directions such as the armchair (AC) and the zigzag (ZZ) direction have been applied. The applied strain is defined as $\epsilon = \frac{\text{a} - \text{a}_0}{\text{a}_0} \times 100\%$, where a$_0$ and a are the unstrained and strained lattice parameters, respectively. The effect of strain on the band gap and the band structure plots under extreme strain cases (AC$\_10\%$ and ZZ$\_10\%$) using the GGA-PBE functional are shown in Fig. \ref{Fig.3}, and that using the HSE06 functional are presented in the supplementary material (see supplementary Sec. A Fig. S1). From Fig. S1, it can be seen that not only the band dispersions but also the magnitude of reduction in the band gap under the strained conditions are well captured within the GGA-PBE XC functional. We, therefore, affirm that the choice of the GGA-PBE functional for the electronic structure calculations of ML-MoS$_2$ is a legitimate one. The band structure plots in the intermediate strain values are shown in Fig. S5 (see supplementary Sec. B Fig. S5). In all strain cases, a direct to indirect band gap transition at low values of strain, followed by a lowering of the band gap is seen. The reduction in the band gap is nearly identical for strains along the AC and the ZZ directions (see Fig. \ref{Fig.3} (b)). It can be seen from Fig. \ref{Fig.3} (a) that the reduction in the band gap is mainly due to the shifting of the conduction band (CB) edge at K towards the Fermi energy. In the unstrained condition, the valence band edges at K and $\Gamma$ are nearly degenerate in energy and thereby form two hole-pockets in the valence band edge (see Fig. \ref{Fig.1} (b)). With tensile strain, the valence band (VB) edge at K starts to move towards lower energy, shifting the VBM from the K-point to the $\Gamma$-point (see Fig. \ref{Fig.3}). Therefore, only the hole-pocket at $\Gamma$ remains at the valence band edge in the strained structures. The magnitude of the energy shifts of the VB and the CB edges (i.e. the VBM and CBM) with strain can be seen in Fig. S6 (see supplementary Sec. B Fig. S6). A detailed analysis of how various in-plane strains on ML-MoS$_2$ can alter the band edge energies has been presented in our previous work \cite{chaudhuri2022strain}. From Fig. S2 in the supplementary material (see supplementary Sec. A Fig. S2) it can be seen that the VB edge at K-point is mainly composed of the in-plane d$_{xy}$/d$_{x^{2}-y^{2}}$ orbitals of Mo, and the VB edge at $\Gamma$ is composed of the hybridized out-of-plane Mo-d$_{z^{2}}$ and S-p$_{z}$ orbitals. The CB edge at the K-point is found to be composed of the Mo-d$_{z^{2}}$ orbitals in anti-bonding configuration. Strain induced changes in the structural parameters, such as the Mo-S bond length and the S-S interplanar distance, result in changes in the hybridization strength between the Mo-d and S-p orbitals and thereby shifting the VB and the CB edges in energy. With the application of tensile strain, the Mo-Mo bond length increases, while a decrease is observed in the S-Mo-S bond angle and the S-S interplanar distance. Therefore, under tensile strain, the interaction strength between the out-of-plane Mo-d$_{z^{2}}$ orbitals is weakened. Thus, with increasing tensile strain, the CB states at K, comprising of the Mo-d$_{z^{2}}$ orbitals in anti-bonding configuration, shift down in energy.

\begin{figure}[h!]
 \centering
 \includegraphics[scale=0.6]{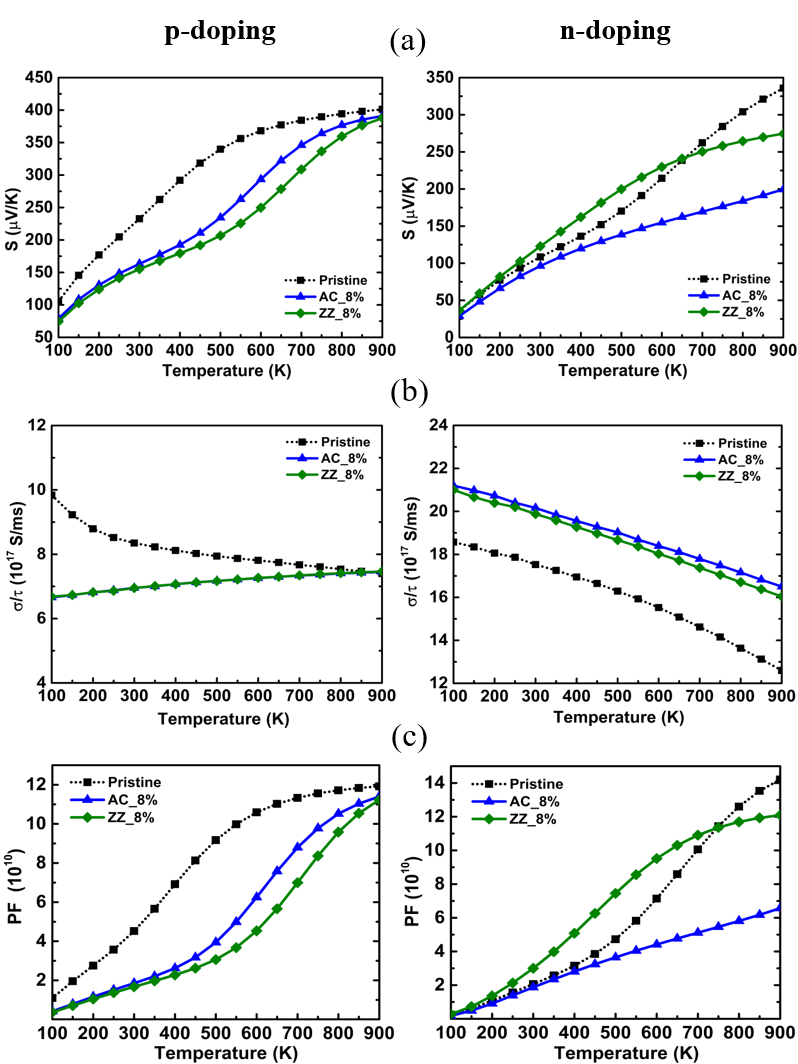}
 \caption{Variation in the (a) Seebeck coefficient (S), (b) electrical conductivity ($\sigma/\tau$), and (c) power factor ($\text{S}^{2}\sigma/\tau$) of ML-MoS$_2$ with temperature at different strain values. The curves for the unstrained structure are shown with the dotted lines. The left panel stands for p-type carriers and the right panel for n-type carriers. The above transport calculations are performed with a fixed doping concentration of $5 \times 10^ {19}$ cm$^{-3}$ for both p- and n-doping.  }
 \label{Fig.4}
\end{figure}

\subsubsection{Effect of strain on thermoelectric properties}
The change in electronic structure is expected to influence the transport and thermoelectric properties. The variation in the thermoelectric parameters (S, $\sigma/\tau$, PF) of single-layer MoS$_2$ with mechanical strain in a temperature range of $100\, \text{K}$ to $900\, \text{K}$ is shown in Fig. \ref{Fig.4}. In all the thermoelectric calculations, the doping level is kept fixed at $5 \times 10^ {19}\, \text{cm}^ {-3}$ for both p- and n-type carriers. Such a doping concentration has been used in earlier theoretical works \cite{bhattacharyya2014effect} and has been achieved also in experiments \cite{hippalgaonkar2017high}. Under the action of tensile strains the value of S is found to decrease. Similar observations have also been reported in earlier works \cite{jena2017compressive, guo2013high, xiang2019monolayer}. For $8\%$ uniaxial strain along the ZZ direction only, the values of S exceed that of the unstrained cases for temperatures up to $700\, \text{K}$, as can be seen from Fig. \ref{Fig.4}(a). At $300\, \text{K}$, the S value drops from $235\, \mu \text{V/K}$ in the unstrained condition to nearly $155\, \mu \text{V/K}$ in the strained cases for p-type carriers. The thermopower in the unstrained, as well as the strained cases, is found to increase with temperature for both p and n doping.  

From the variation of the relaxation-time-coupled electrical conductivity ($\sigma/\tau$) with strain and temperature (see Fig. \ref{Fig.4}(b)), it can be seen that the $\sigma/\tau$ varies differently for p- and n-type carriers under the action of strain. For p-type carriers, the $\sigma/\tau$ decreases with strain, whereas it increases for n-type carriers. Also, the values of $\sigma/\tau$ are higher for n-doping than that of the p-doping for unstrained and strained cases throughout the temperature range $100\, \text{K}$ to $900\, \text{K}$. This contrasting behaviour of the p- and n-type carriers can be understood from the dispersion of the valence and conduction band edges of ML-MoS$_2$. With strain application, the VBM shifts from K-point to $\Gamma$-point, where the dispersion is significantly less, whereas the CBM at K-point becomes more dispersive compared to the unstrained condition. The same can be seen from the calculated effective masses associated with the band edges, as discussed later. Due to this highly dispersive nature, the CBM offers very high mobility of the charge carriers. Therefore, the $\sigma/\tau$ of the n-type carriers becomes higher compared to the p-type carriers. The temperature variation of the $\sigma/\tau$ also shows different characteristics for n- and p-type carriers. The changes in $\sigma/\tau$ with temperature remain insignificant for p-type doping, whereas it decreases gradually for n-type doping. This is probably due to the intrinsic n-type nature of monolayer MoS$_2$ in the pristine condition. With the further addition of n-type carriers, it starts to behave like a degenerate semiconductor and hence, the conductivity falls with increasing temperature. However, to estimate the actual variation in $\sigma$ with strain and temperature, decoupling of the carrier relaxation time ($\tau$) from $\sigma/\tau$ is necessary.     

The usefulness of a thermoelectric material is gauged by its power factor ($\text{S}^{2}\sigma$). The power factor (PF) should be high for a good thermoelectric material. The variation in the relaxation-time-scaled PF ($\text{S}^{2}\sigma/\tau$) of monolayer MoS$_2$ with strain and temperature is shown in Fig. \ref{Fig.4}(c). It can be seen that, except for the case of n-doped $8\%$ uniaxial tension along the ZZ direction, the PF value decreases with tensile strain compared to the pristine values throughout the temperature range. In the unstrained condition, owing to the presence of two nearly degenerate hole-pockets in the valence band (at K and $\Gamma$), the PF values obtained with p-doping ($\mu < 0$) are higher than that with n-doping ($\mu > 0$). At room temperature, for the unstrained structure, the PF value obtained with p-doping ($4.51 \times 10^{10}\, \text{Wm}^{-1}\text{K}^{-2}\text{s}^{-1}$) is twice as high as that with n-doping ($2.05 \times 10^{10}\, \text{Wm}^{-1}\text{K}^{-2}\text{s}^{-1}$). With increasing strain, the hole-pocket at K moves downward in energy. Therefore, only the valence band states at $\Gamma$, which has much less dispersion, contribute to carrier transport. Owing to the 6-fold degeneracy of the valence band states at K, the density of states effective mass (m$_\text{D}$) is very high in the unstrained condition. However, in the strained cases, the m$_\text{D}$ is significantly reduced due to the non-degenerate $\Gamma$ point. Thus, the power factor falls dramatically with strain for p-type carriers. However, the PF value increases gradually with increasing temperature for both types of carriers. At $900\, \text{K}$, the highest PF value of $14.3 \times 10^{10}\, \text{Wm}^{-1}\text{K}^{-2}\text{s}^{-1}$ is obtained with the n-doped unstrained structure, which is higher compared to the values obtained with the strained structures. \iffalse The efficiency (ZT) of a thermoelectric material is proportional to its power factor. Thus, the reduction in power factor under the action of strain indicates that tensile strain has a diminishing effect on the thermoelectric performance of ML-MoS$_2$. However, the application of in-plane tensile strain simultaneously reduces the lattice thermal conductivity ($\kappa_\text{L}$), to which the thermoelectric efficiency (ZT) is inversely proportional. Therefore, a large reduction in $\kappa_\text{L}$ may in turn increase the thermoelectric efficiency of single-layer MoS$_2$. \fi

\begin{figure}[h!]
	\centering
	\includegraphics[scale=0.5]{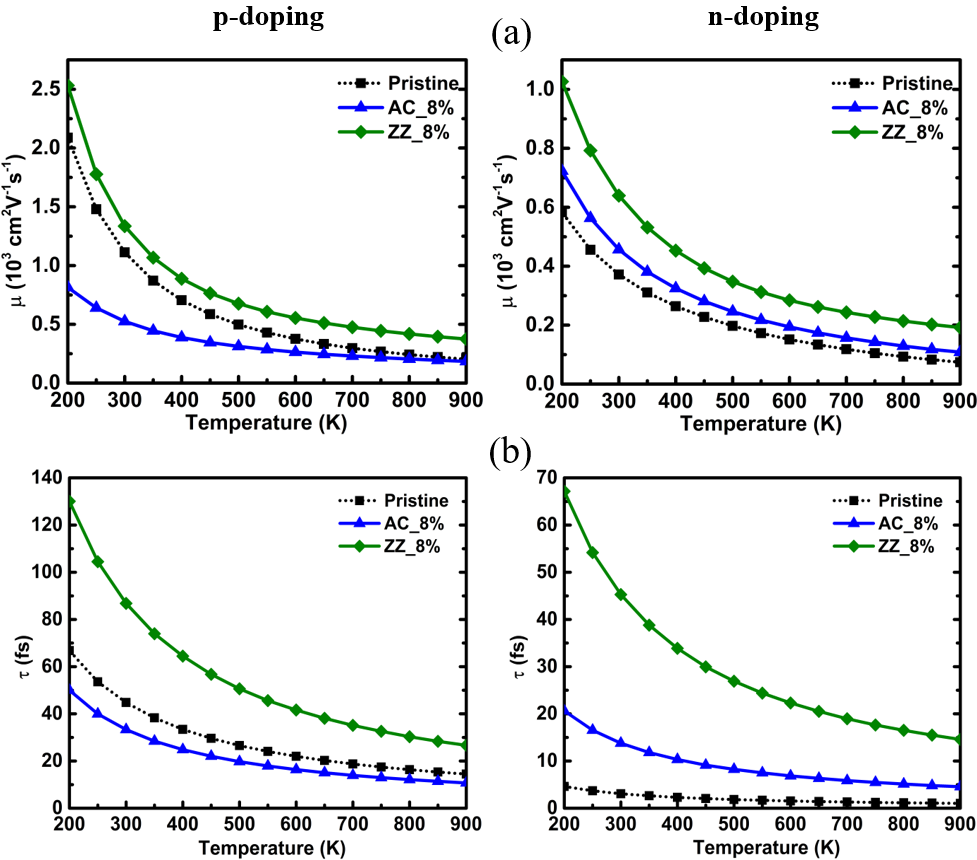}
	\caption{Variation in the (a) mobility ($\mu$), (b) relaxation time ($\tau$) of the charge carriers of ML-MoS$_2$ with temperature at different strain values. The curves for the unstrained structure are shown with the dotted lines. The left panel stands for holes and the right panel for electrons.}
	\label{Fig.5}
\end{figure}

It is worth noting that the thermoelectric parameters calculated herein, such as the $\sigma$, $\kappa_\text{e}$, $\text{S}^{2}\sigma$ are scaled by the electronic relaxation time ($\tau$), since all calculations are performed based on the constant relaxation time approximation (CRTA). The CRTA assumes that the electronic relaxation time does not vary strongly with energy. Such an oversimplified assumption, at times, leads to conclusions far from reality. In the present case, the explicit determination of $\tau$ is essential, due to the fact that the carrier relaxation time itself is a function of strain. In the calculations of carrier mobility ($\mu$) and relaxation time ($\tau$), the scattering of the charge carriers with phonons i.e., with longitudinal acoustic phonons within the deformation potential theory (ADP) and with polar optical phonons (POP) are considered. Other scattering processes, such as the boundary scattering or ionized impurity scattering may also play a vital role in dictating the actual scattering lifetimes observed in typical experiments. However, they are strongly subjected to experimental parameters and thus, have been ignored henceforth. Due to the non-consideration of these scattering processes, the calculations are expected to provide only an intrinsic limit of the $\mu$ and $\tau$, which can be largely overestimated compared to the experimental values. The total mobility of the charge carriers is calculated using the Matthiessen's rule ${\mu}^{-1} = \sum {\mu_{i}^{-1}}$, where $i$ stands for different scattering events, such as ADP and POP. The variation in the total mobility and relaxation time of electrons and holes with strain and temperature are shown in Fig. \ref{Fig.5}. The individual contribution of the different scattering processes, such as the ADP and POP, towards the total $\mu$ and $\tau$, are presented in Fig. S7 and S8 (see supplementary information Sec. B Fig. S7 and S8). It can be seen that the charge carrier transport is mainly governed by the ADP and the contribution from the POP is negligibly small. The inconsequential effect of the POP can be understood due to the intrinsically non-polar nature of MoS$_{2}$. Therefore, it is legitimate to focus only on the ADP to understand the effect of strain and temperature on the carrier transport.

\begin{table}[h]
	\caption{\label{tab:Table 2} The calculated physical parameters i.e., the effective mass ($m^{*}$) of electrons and holes, the effective 2D elastic constant ($C_\text{2D}$) in GPa, and the deformation potential constant ($E_\text{DP}$) for the conduction and the valence band edges in eV of ML-MoS$_2$ in the unstrained and $8\%$ strained cases along the AC and the ZZ directions.}
	\begin{tabular*}{0.9\textwidth}{ c @{\extracolsep{\fill}} c c c c c c c}
		\hline
		&  & $m^{*}$ & & $C_\text{2D}$ & & $E_\text{DP}$ & \\
		\hline
		&  & AC & ZZ & AC & ZZ & AC & ZZ \\
		\hline 
		Pristine & \multicolumn{1}{p{1cm}}{\centering e \\h} & \multicolumn{1}{p{2cm}}{\centering 0.46 \\0.56} & \multicolumn{1}{p{2cm}}{\centering 0.47 \\0.59} & \multicolumn{1}{p{2cm}}{\centering 215 \\215} & \multicolumn{1}{p{2cm}}{\centering 215 \\215} & \multicolumn{1}{p{2cm}}{\centering -6.35 \\-2.54} & \multicolumn{1}{p{2cm}}{\centering -6.67 \\-2.31}  \\
		\hline  
		AC$\_{8\%}$ & \multicolumn{1}{p{1cm}}{\centering e \\h} & \multicolumn{1}{p{2cm}}{\centering 0.42 \\2.5} & \multicolumn{1}{p{2cm}}{\centering 0.40 \\2.17} & \multicolumn{1}{p{2cm}}{\centering 149 \\149} & \multicolumn{1}{p{2cm}}{\centering 207 \\207} & \multicolumn{1}{p{2cm}}{\centering -5.43 \\-2.31} & \multicolumn{1}{p{2cm}}{\centering -5.96 \\-2.01} \\
		\hline
		ZZ$\_{8\%}$ & \multicolumn{1}{p{1cm}}{\centering e \\h} & \multicolumn{1}{p{2cm}}{\centering 0.43 \\1.32} & \multicolumn{1}{p{2cm}}{\centering 0.39 \\1.27} & \multicolumn{1}{p{2cm}}{\centering 193 \\193} & \multicolumn{1}{p{2cm}}{\centering 137 \\137} & \multicolumn{1}{p{2cm}}{\centering -5.38 \\-1.12} & \multicolumn{1}{p{2cm}}{\centering -5.10 \\-0.98} \\
		\hline
	\end{tabular*}
\end{table}

The variation in the acoustic phonon limited (ADP) carrier transport properties with strain and temperature can be understood using the simplest form of the deformation potential theory of Bardeen and Shockley \cite{bardeen1950deformation}, where the carrier mobility is given as,	 $\mu = \frac{2e\hbar^{3}C_{2D}}{3k_{B}T{m^{*}}^{2}(E_\text{DP})^{2}}$, where $C_\text{2D}$ is the effective elastic constant, $m^{*}$ is the carrier effective mass, and $E_\text{DP}$ is the deformation potential constant of the VBM for the holes and CBM for the electrons. Note that, the original form of the Bardeen-Shockley equation, as shown here for the sake of simplicity, does not consider the influence of transverse acoustic phonon modes in the calculation of carrier mobility. However, in this work, the effect of transverse phonons and thereby, the anisotropy in the $C_\text{2D}$, $m^{*}$ and $E_\text{DP}$ have been taken into account in the calculation of deformation potential limited carrier mobility. The calculated elastic constants, effective mass and the deformation potential constants for the unstrained and strained cases are provided in Table \ref{tab:Table 2} and they are in good agreement with earlier reports \cite{wiktor2016absolute, hung2018two, laturia2018dielectric}. The carrier mobility decreases as a function of temperature following the inverse law, as can be seen from Fig. \ref{Fig.5}. The mobility of the electrons and holes at $300$\, K are found to be $368$ and $1100$ $\text{cm}^{2}\text{V}^{-1}\text{s}^{-1}$, respectively. Due to the small deformation potential of the VBM, the hole mobilities are found to be much higher compared to the electron mobilities. The room temperature mobilities calculated herein are seemingly higher as compared to the earlier reports \cite{rawat2018comprehensive, ding2021geometry}. The observed large difference is primarily due to two important factors, one is the consideration of carrier screening in our calculations and the other one is the accurate determination of the absolute deformation potential constant. Since all the electronic transport calculations are done considering a high carrier concentration ($5 \times 10^ {19}\, \text{cm}^ {-3}$), it is expected that the large carrier density would impose significant screening and thereby impact the transport properties. Next, the deformation potential constants ($E_\text{DP}$) calculated in the earlier studies ($10.7$, $5.1$) are much higher compared to our calculations ($6.6$, $2.3$), due to which the carrier mobilities found therein are much smaller. In the calculation of $E_\text{DP}$, it is necessary to compute the changes in the VBM and CBM energies with respect to a fixed reference level. In our calculations, the vacuum energy level is taken as the reference and the values of $E_\text{DP}$ calculated are found to be in good agreement with the absolute deformation potentials calculated by Pasquarello et al. \cite{wiktor2016absolute}. To get a comparable estimate with the experimental values, we also calculate the electron mobility of unstrained ML-MoS$_2$ considering the scattering of the charge carriers with impurity defects in addition to the intrinsic carrier-phonon scattering and presented in Fig. S9 (see supplementary information Sec. B Fig. S9). With the incorporation of the impurity scattering, the $\mu$ of electrons at $300$\, K are found to be $81.1$ $\text{cm}^{2}\text{V}^{-1}\text{s}^{-1}$, which is in good agreement with previous experimental and theoretical investigations \cite{rawat2018comprehensive, ding2021geometry, zhang2012ambipolar}. Therefore, an overestimated value of $\mu$ and $\tau$ is obtained when only intrinsic carrier-phonon scattering is taken into consideration. However, we stress that the effect of strain and temperature on the carrier transport, which is the aim of this study, can be well captured within this simplified formulation.  

Compared to the carrier mobilities in the unstrained condition, the electron mobilities (${\mu}_e$) increase with the application of tensile strain along both the AC and ZZ directions, whereas the hole mobilities increase only for strain along the ZZ direction. With tensile strain, the effective mass of electrons at the CBM (at K point) decreases monotonically with a large decrease in $E_\text{DP}$ of the conduction band edge, as can be seen from Table \ref{tab:Table 2}. Therefore, the electron mobility (${\mu}_e$) is enhanced with tensile strains at all temperatures from $200$\, K to $900$\, K. The magnitude of enhancement in ${\mu}_e$ is higher for strains along the ZZ direction than the AC direction, which can be understood from the larger reductions in m$^{*}$ and $E_\text{DP}$ associated with it. The variation in the hole mobility with strain is complex due to the shift of the VBM from the high-disperse K point to the much less-disperse $\Gamma$ point. The large increase in the hole effective mass (see Table \ref{tab:Table 2}) resulting from the flat valence band top at $\Gamma$ governs the hole transport in the strained conditions and thus, results in a large decrease in hole mobility. However, for strains along the ZZ direction, the large reduction in $E_\text{DP}$ ($\sim$ 2.5 times for $8\%$ strain) overshadows the impact of large effective mass and results in higher mobility compared to the pristine cases at all temperatures. The carrier relaxation time ($\tau$) for electrons and holes, shown in Fig. \ref{Fig.5}, is computed from the sum of the individual scattering rates ($1/\tau$). The strain-induced variation in the individual scattering rates corresponding to ADP and POP are presented in Fig. S8 (see supplementary information Sec. B Fig. S8). The estimated $\tau$ values are incorporated in the calculated electronic transport parameters such as the $\sigma/\tau$, $\kappa_\text{e}/\tau$, $\text{S}^{2}\sigma/\tau$ to get rid of the relaxation time scaling resulting from the CRTA.                   

\begin{figure}[h!]
 \centering
 \includegraphics[scale=0.4]{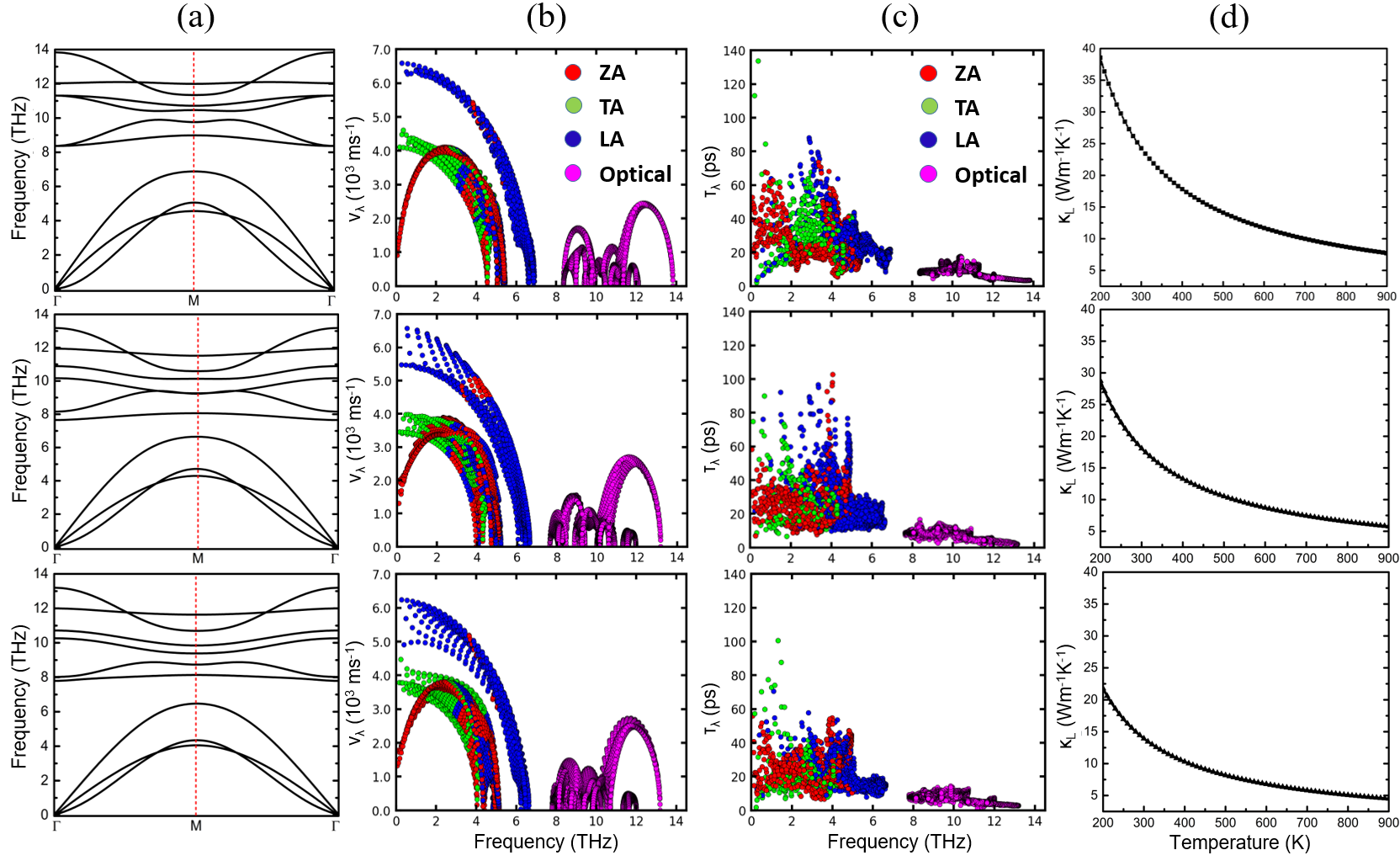}
 \caption{(a) Phonon band structures plotted along the high symmetry path, (b) phonon group velocity ($\text{v}_\lambda$) plotted as a function of frequency (c) relaxation time ($\tau_\lambda$) plotted as a function of frequency at $300$ K, and (d) lattice thermal conductivity ($\kappa_\text{L}$) plotted as a function of temperature for the unstrained (top panel), $8$\% uniaxially strained structures along the armchair (middle panel) and the zigzag (bottom panel) direction.}
 \label{Fig.6}
\end{figure}

Next, the impact of tensile strain on the lattice thermal conductivity ($\kappa_\text{L}$) is investigated, and the variation in $\kappa_\text{L}$ with strain and temperature is presented in Fig. \ref{Fig.6}. For unstrained monolayer MoS$_2$, the value of $\kappa_\text{L}$ at $300\, \text{K}$ is found to be $24.28\, \text{Wm}^{-1}\text{K}^{-1}$, which agrees well with earlier reports \cite{rai2020electronic, cai2014lattice, yan2014thermal}. With in-plane tensile strains, the value of $\kappa_\text{L}$ decreases throughout the temperature range. However, the rate of decrease strictly depends on the direction of the applied strain.  The reduction in $\kappa_\text{L}$ is found to be higher for uniaxial strains along the zigzag (ZZ) direction than along the armchair (AC) direction for all temperatures (see Fig. \ref{Fig.6} (d)). The values of $\kappa_\text{L}$ at $300\, \text{K}$ for $8\%$ uniaxial strain along the AC and the ZZ direction are found to be $18\, \text{Wm}^{-1}\text{K}^{-1}$ and $14\, \text{Wm}^{-1}\text{K}^{-1}$ respectively, compared to the pristine value of $24.28\, \text{Wm}^{-1}\text{K}^{-1}$. Such anisotropic reduction in $\kappa_\text{L}$ with strain can be understood from the difference in atomic coordination along the two non-equivalent directions, where unique nearest neighbour (NN) patterns can be seen. Each S atom at the edge possesses two NN Mo atoms, each of which has three NN S atoms along the ZZ direction, while along the AC direction, one Mo atom has two and the other one has three NN S atoms. Due to this difference in nearest neighbour counting, the applied strain of equal magnitude has a different impact on $\kappa_\text{L}$ along the two directions (AC and ZZ). Also, the $\kappa_\text{L}$ values decrease gradually with increasing temperature following the $\dfrac{1}{\text{T}}$ law owing to the increased probability of Umklapp scattering and becoming less than $5\, \text{Wm}^{-1}\text{K}^{-1}$ at $900\, \text{K}$ for strains applied along the ZZ direction. The total thermal conductivity computed by adding the electronic ($\kappa_\text{e}$) and the lattice contribution ($\kappa_\text{L}$), is provided in Fig. S10 (see supplementary Sec. C Fig. S10). At all strain values and temperatures, the total thermal conductivity is dominated by the lattice counterpart. The electronic contribution is at least two orders of magnitude lower than the lattice contribution, as shown in Fig. S10 (see supplementary Sec. C Fig. S10). This can be understood as the calculations are done within the semiconductor regime of ML-MoS$_2$, and for a semiconductor, the majority of the heat is carried by the phonons with negligible contribution stemming from the electrons.

To investigate the strain-induced large and anisotropic reduction in $\kappa_\text{L}$ in greater detail, further analysis of the $\kappa_\text{L}$ at $300\, \text{K}$ is performed. From the cumulative lattice thermal conductivity ($\kappa_{\text{L}_\text{c}}$) (see supplementary Sec. C Fig. S11), it is clear that almost the entire contribution towards $\kappa_\text{L}$ is coming from the frequencies below $7$ THz i.e., from the acoustic phonon modes, as expected. The contribution coming from the optical modes is negligible. To investigate the contribution from individual acoustic modes, modal-decomposed values of $\kappa_\text{L}$ and various other parameters that help in the analytical modelling of $\kappa_\text{L}$ such as the specific heat ($\text{C}_\lambda$), group velocity ($\text{v}_\lambda$) and relaxation time ($\tau_\lambda$) of a particular phonon mode $\lambda$ are computed at $300\, \text{K}$ and are presented with the phonon band structures in Fig. \ref{Fig.6}. From the mode resolved $\kappa_\text{L}$ shown in Fig. S12 (see supplementary Sec. C Fig. S12), it is clear that the contribution stemming from the three acoustic modes is reduced under the action of tensile strains. Both the $\text{v}_\lambda$ and $\tau_\lambda$ of the acoustic modes are considerably higher compared to the optical modes and thus, the acoustic modes dominate the heat transport mechanism. With increasing strain, both $\text{v}_\lambda$ and $\tau_\lambda$ decrease, although the reduction in $\tau_\lambda$ is found to be more pronounced. Also, the reduction in $\tau_\lambda$ for strains along the ZZ direction is higher than that for the AC direction. Due to the reduction in $\tau_\lambda$, the phonon mean free path reduces and the phonon scattering probability increases significantly. As a result, the $\kappa_\text{L}$ decreases with strain and a larger degree of reduction occurs for strain along the ZZ direction. In MoS$_2$, as well as other $2$D-TMDCs, the ZA mode is seen to be the dominant heat conduction carrier and the frequency gap between the ZA and the optical modes determines the intensity of the phonon scattering \cite{wang2021improved, peng2019dominant}. The smaller the frequency gap, the stronger the scattering and the lower the $\kappa_\text{L}$. In an earlier report \cite{chaudhuri2022strain}, it has been observed that with in-plane tensile strains, the ZA-optical frequency gap of ML-MoS$_2$ decreases. This reduction in frequency gap may result in an enhanced ZA-optical phonon scattering and thereby, the $\kappa_\text{L}$ of ML-MoS$_2$ is further lowered under the action of tensile strains.

\begin{table}
	\caption{\label{tab:Table 3} Phonon parameters of single-layer MoS$_2$ in the unstrained and $8\%$ strained conditions along the AC (AC$\_8\%$) and the ZZ (ZZ$\_8\%$) directions for the three acoustic modes (labelled according to polarization): group velocity (v$_{\lambda}$) at $\Gamma$ point in $10^{3}$ ms$^{-1}$ and acoustic Debye temperature ($\theta_{\text{D}}$) in K.}
	
	\begin{tabular*}{1.0\textwidth}{ c @{\extracolsep{\fill}} c c c c c c c}
		\hline
		Strain profile & $\text{v}_{\lambda \_ \text{ZA}}$ & $\text{v}_{\lambda \_ \text{TA}}$ & $\text{v}_{\lambda \_ \text{LA}}$ & $\theta_{\text{D} \_ \text{ZA}}$ & $\theta_{\text{D} \_ \text{TA}}$ & $\theta_{\text{D} \_ \text{LA}}$ \\
		\hline 
		Unstrained & 0.81 & 4.38 & 6.46 & 246 & 223 & 334 \\
		\hline  
		AC\_8\% & 1.27 & 3.61 & 5.7 & 230 & 209 & 322 \\
		\hline  
		ZZ\_8\% & 1.27 & 3.75 & 6.15 & 212 & 197 & 314 \\
		\hline 
	\end{tabular*}
\end{table}

Furthermore, the high $\kappa_\text{L}$ of pristine ML-MoS$_2$ results from the low average atomic mass and the strong in-plane covalent bonding between the Mo and the S atoms. Strong interatomic bonding and low average atomic mass results in larger Debye temperature ($\theta_{\text{D}}$) and higher sound velocity, which in turn results in higher lattice thermal conductivity. The modal group velocity ($\text{v}_{\lambda}$) near the zone centre ($\Gamma$) and the Debye temperature ($\theta_\text{D}$) at the zone boundary of the acoustic phonon modes are listed in Table \ref{tab:Table 3}, which are in good agreement with previous reports \cite{peng2016thermal}. It can be seen that both the v$_\lambda$ and the $\theta_{\text{D}}$ decrease with the application of strain and thus, result in the lowering of the $\kappa_\text{L}$ according to Slack's expression \cite{slack1973nonmetallic}. With tensile strain, the orbital overlap of the Mo and S atoms decreases and therefore, the interatomic bond becomes weaker. Thus, the Debye temperature decreases with increasing tensile strain. The reduction in $\theta_{\text{D}}$ is higher when the strain is applied along the zigzag direction (see Table \ref{tab:Table 2}) and thereby results in a greater reduction of $\kappa_\text{L}$ compared to that of the armchair direction. The strain-induced changes in the acoustic $\text{v}_{\lambda}$ are not significant and are mostly shadowed by the large changes in the $\tau_{\lambda}$. Also, the anharmonicity or the Gr$\ddot{\text{u}}$neisen parameter of a material controls the ability of heat transportation through the crystal lattice. The modal Gr$\ddot{\text{u}}$neisen parameter of ML-MoS$_2$ is calculated in order to understand the degree of anharmonicity of the material. No significant change in the Gr$\ddot{\text{u}}$neisen parameter is observed in the strained cases compared to the unstrained values (see supplementary sec. C Fig. S13). Thus, the strain-induced anisotropic reduction in $\kappa_\text{L}$ is primarily due to the anisotropic changes in the bond parameters and the resulting $\theta_{\text{D}}$ of the acoustic phonon modes.

\begin{figure}[h!]
 \centering
 \includegraphics[scale=0.4]{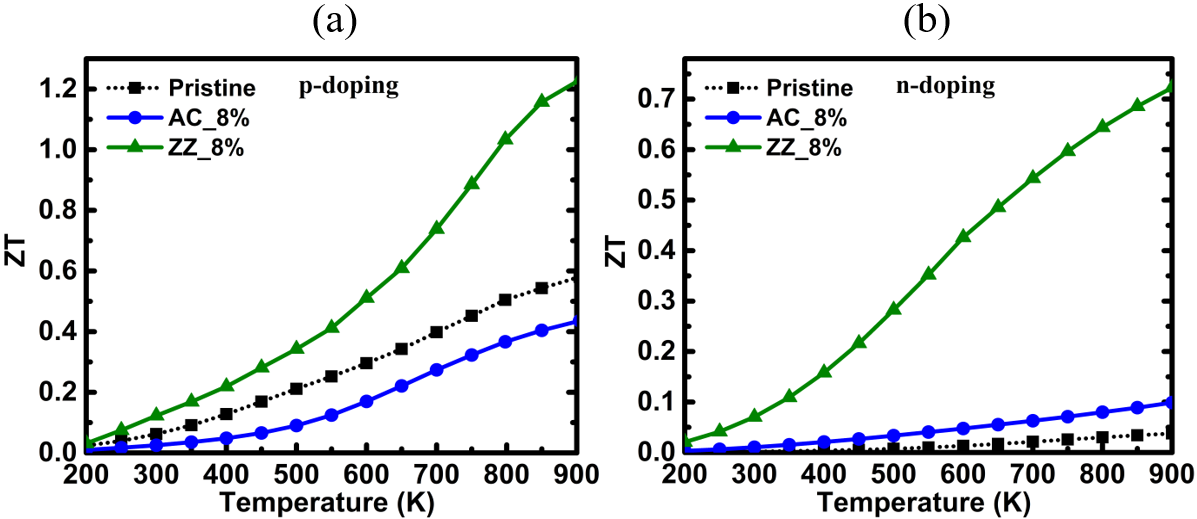}
 \caption{The variation in thermoelectric figure of merit ZT of ML-MoS$_2$ with temperature at different tensile strains for (a) p-doping and (b) n-doping.}
 \label{Fig.7}
\end{figure}

Finally, combining all the calculated transport parameters (S, $\sigma$, $\tau$, $\kappa$), the thermoelectric efficiency of ML-MoS$_2$ is estimated and its variation with chemical potential ($\mu$) and temperature (T) in the unstrained condition is explored (see supplementary Sec. C Fig. S14). The results show that the maximum value of ZT for pristine MoS$_2$ at $300\, \text{K}$ for both p-type and n-type doping is less than $0.1$. Such a low value of ZT is not useful for practical applications. However, the ZT value is found to increase sharply with temperature and reach close to $0.7$ at $900\, \text{K}$ for p-type carriers, suggesting that ML-MoS$_2$ can be an ideal candidate for high temperature thermoelectric applications.

The impact of strain and temperature on the ZT of ML-MoS$_2$ at a fixed doping concentration of $5 \times 10^{19}\, \text{cm}^{-3}$ is investigated and the results are displayed in Fig. \ref{Fig.7}(a) and (b). For both p- and n-doping, the ZT values of the strained cases exceed that of the unstrained values throughout the temperature range and higher ZT values can be obtained when the strain is applied along the ZZ direction. The peak ZT values obtained at $900\, \text{K}$ for p- and n-doping with $8\%$ uniaxial strain along the ZZ direction are $1.22$ and $0.72$, respectively. These ZT values are significantly higher compared to the peak values obtained with the unstrained structure ($0.58$ and $0.03$) and the $8\%$ strained structure along the AC direction ($0.43$ and $0.09$). The larger enhancement in charge carrier relaxation time and greater reduction in $\kappa_{\text{L}}$ for strains along the ZZ direction act in unison to result in the higher ZT values compared to that for strains along the AC direction. From the larger values of ZT for holes and its variation with strain, it is clear that the charge carrier relaxation time ($\tau$) has significant control on the overall thermoelectric performance of ML-MoS$_2$. Therefore, it is wise to go beyond CRTA, and the explicit determination of $\tau$ is necessary. It is worth noting that the enhanced thermoelectric performance is not only achieved at the specific doping concentration of $5 \times 10^{19}\, \text{cm}^{-3}$ or a high strain value of $8\%$; instead, the enhancement can be achieved for a range of doping concentrations and even with lower strain values. To further support this point, the variation in ZT at $900\, \text{K}$ as a function of chemical potential under $4\%$ and $8\%$ uniaxial tensile strains along the AC and ZZ directions are calculated and shown in Fig. S15 (see supplementary Sec. C Fig. S15). It is clear that a much-improved thermoelectric performance can be achieved with the strained structures at the optimal doping concentration. The degree of enhancement, though, strictly depends on the direction and magnitude of the strain applied. The ZT values obtained with the strained cases exceed that of the unstrained case throughout the range of chemical potential or the corresponding doping concentrations. A peak ZT value of $1.7$ can be achieved with the hole-doped ML-MoS$_2$ under $8\%$ uniaxial strain along the zigzag direction. Compared to the peak ZT value of $0.7$ obtained with the hole-doped unstrained structure, an enhancement of nearly $150 \%$ is possible with the strained structures.

\section{Conclusions}
In summary, the anisotropic tuning of the transport and thermoelectric properties of monolayer MoS$_2$ with the application of in-plane tensile strains along the armchair (AC) and the zigzag (ZZ) direction has been explored based on first-principles calculations. Both the electronic and phononic transport properties of single-layer MoS$_2$ change anisotropically with the application of direction-dependent mechanical strains, suggesting that strain engineering can be an efficient way to tune the thermoelectric properties of ML-MoS$_2$. The Seebeck coefficient (S) and the power factor ($\text{S}^{2}\sigma/\tau$) of ML-MoS$_2$ decrease with tensile strains for both p- and n-type carriers. However, the reduction is more pronounced for p-type carriers. This is understood from the strain-induced reduction in the number of degenerate hole-pockets in the valence band edge. Although the power factor decreases, the thermoelectric efficiency of ML-MoS$_2$ is found to increase due to the large reduction in lattice thermal conductivity ($\kappa_\text{L}$) and the significant increase in charge carrier relaxation time ($\tau$) with tensile strain. Notably, the reduction in $\kappa_\text{L}$ and increase in $\tau$ associated with the tensile strains along the ZZ direction are significantly higher compared to the strains along the AC direction. Therefore, a large enhancement in ZT is achieved with the structures strained along the ZZ direction. With optimal doping at $900\, \text{K}$, a ZT value as high as $1.7$ is achieved with the strained structures compared to the peak ZT value of $0.7$ in the unstrained structure and structures strained along the AC direction. This study, therefore, highlights the importance of direction-specific tensile strains in improving the overall transport characteristics of ML-MoS$_2$. The proposed mechanism to enhance thermoelectric performance is expected to be equally useful for all other semiconducting TMDCs with an analogous crystal structure. 

\begin{acknowledgments}
	The first-principles calculations are performed using the supercomputing facility of IIT Kharagpur established under the National Supercomputing Mission (NSM), Government of India and supported by the Centre for Development of Advanced Computing (CDAC), Pune. AB acknowledges SERB POWER grant (SPG/2021/003874) and BRNS regular grant (BRNS/37098) for the financial assistance. SC acknowledges MHRD, India, for financial support.
\end{acknowledgments}

\bibliographystyle{achemso}
\bibliography{biblio}

\end{document}